\documentclass[aps,prl,amsfonts,amsmath,amssymb,reprint,twocolumn,superscriptaddress,showpacs,a4paper]{revtex4-1}

\usepackage{color}
\usepackage{graphicx}
\usepackage{hyperref}
\usepackage{siunitx}
\usepackage{bbm}

\usepackage{ulem}

\newcommand{\rb}{$^{87}\mathrm{Rb}\ $}
\newcommand{\var}{\mathrm{var}}
\newcommand{\ave}[1]{\ensuremath{\langle#1\rangle}}
\newcommand{\NA}{	N_{\rm{A}}}
\newcommand{\NL}{N_{\rm{L}}}
\newcommand{\Npulse}{N_{\rm{p}}}

\newcommand{\sz}{S_{\rm z}}

\newcommand{\Bz}{B_{\rm z}}
\newcommand{\By}{B_{\rm y}}
\newcommand{\Bx}{B_{\rm x}}

\newcommand{\Larmor}{\omega_{L}}

\newcommand{\fx}{F_{\rm x}}
\newcommand{\fy}{F_{\rm y}}
\newcommand{\fz}{F_{\rm z}}

\newcommand{\Fx}{F_{\rm x}}
\newcommand{\Fy}{F_{\rm y}}
\newcommand{\Fz}{F_{\rm z}}
\newcommand{\bF}{{\bf F}}

\newcommand{\RF}{RF}

\newcommand{\app}{\hat{a}_{+}}
\newcommand{\am}{\hat{a}_{-}}
\newcommand{\apd}{\hat{a}_{+}^{\dagger}}
\newcommand{\amd}{\hat{a}_{-}^{\dagger}}
\newcommand{\apm}{\hat{a}_{\pm}}
\newcommand{\sigmai}{\sigma_i}

\newcommand{\bB}{\bf{B}}

\newcommand{\Mone}{\mathcal{M}_1}
\newcommand{\Mtwo}{\mathcal{M}_2}

\newcommand{\nuProbe}{\nu_{\rm{probe}}}
\newcommand{\Thold}{{t_{\rm hold}}}

\newcommand{\Ttwo}{\rm{T}_2}

\newcommand{\Syin}{S_{\rm y}^{\rm{(in)}}}
\newcommand{\Syout}{S_{\rm y}^{\rm{(out)}}}
\newcommand{\Sxin}{S_{\rm x}^{\rm{(in)}}}

\newcommand{\ff}{f^{(i)}}
\newcommand{\bss}{\bf S}
\newcommand{\Sx}{S_{\rm x}}
\newcommand{\Sy}{S_{\rm y}}
\newcommand{\Sz}{S_{\rm z}}
\newcommand{\Si}{S_{\rm i}}

\newcommand{\probePer}{\nu_{\rm{probe}}}

\definecolor{mygreen}{rgb}{0,0.5,0} 
\definecolor{myblue}{rgb}{0,0,0.75} 
\definecolor{myyellow}{rgb}{0.87,0.8,0.47} 
\definecolor{mymagenta}{cmyk}{0,1,0,0.12} 
\definecolor{myorange}{rgb}{.95,0.5,0}

\newcommand{\supone}{^{(1)}}
\newcommand{\suptwo}{^{(2)}}

\newcommand{\Tend}{t_{\rm end}}

\newcommand{\PRLsection}[1]{\noindent\textit{#1} --}

\begin{document}

\title{
Entanglement-enhanced radio-frequency field detection and waveform sensing \\
}

\newcommand{\ICFOAddress}{ICFO-Institut de Ciencies Fotoniques, The Barcelona Institute of Science and Technology, 08860 Castelldefels (Barcelona), Spain}

\newcommand{\ICREAAddress}{ICREA -- Instituci\'{o} Catalana de Re{c}erca i Estudis Avan\c{c}ats, 08015 Barcelona, Spain}

\newcommand{\OlomoucAddress}{Department of Optics, Palack\'{y} University, 17. listopadu 1192/12, 771 46 Olomouc, Czech Republic}

\author{F. Martin Ciurana}
\affiliation{\ICFOAddress}

\author{G. Colangelo}
\affiliation{\ICFOAddress}

\author{L. Slodi\v{c}ka}
\affiliation{\OlomoucAddress}

\author{R.~J.~Sewell}
\affiliation{\ICFOAddress}

\author{M.~W.~Mitchell}
\affiliation{\ICFOAddress}
\affiliation{\ICREAAddress}

\date{\today}

\newcommand{\SuppInfo}{Supplementary Information}
\newcommand{\Ttot}{T_{\rm acq}}
\newcommand{\Tcoh}{\tau}

\begin{abstract}
We demonstrate a new technique for detecting components of arbitrarily-shaped radio-frequency waveforms based on stroboscopic back-action evading measurements. 
We combine quantum non-demolition measurements and stroboscopic probing to detect waveform components with magnetic sensitivity beyond the standard quantum limit.
Using an ensemble of $1.5\times 10^6$ cold rubidium atoms, we demonstrate entanglement-enhanced sensing of sinusoidal and linearly chirped waveforms, with \SI{1.0 (2)}{dB} and \SI{0.8 (3)}{dB} metrologically relevant noise reduction, respectively. We achieve volume-adjusted sensitivity of $\delta\rm{B}\sqrt{V}\approx 11.20~\rm{fT\sqrt{cm^3/Hz}}$, comparable to the best \RF~magnetometers.
\end{abstract}

\maketitle

Quantum noise and quantum coherence both play essential roles in determining the fundamental sensitivity of interferometric instruments such as atomic magnetometers and atomic clocks. 
This is clearly seen in a widely-used rule of thumb for the projection noise limit $\delta B_{\rm PN}$ of sensing magnetic fields with $\NA$ spin-$f$ atoms~\cite{BudkerNP2007}: 
\begin{equation}
\label{eq:sensitivity}
\delta B_{\rm PN} \sqrt{\Ttot} \simeq \frac{\hbar}{g \mu_B}\frac{1}{\sqrt{2 f \NA\Tcoh}}.
\end{equation}
Here $\Ttot$ is the total acquisition time including averaging repeated measurements, $\mu_{\rm B}$ is the Bohr magneton, $g$ is the ground-state Land\'{e} factor, and $\hbar$ is Planck's constant. 
The factor $1/\sqrt{2 f \NA \Tcoh}$ reflects the standard quantum limit (SQL) spin projection noise of the atomic precession angle, which scales as $1/\sqrt{\NA}$.
The signal accumulation time $\Tcoh$ is determined by the smaller of the single-measurement duration and spin atomic coherence time. Analogous expressions govern clocks and other atomic instruments. 

Reduction of spin projection noise below the SQL~\cite{TakanoPRL2009, AppelPNAS2009, SchleierPRL2010, ChenPRL2011, SewellPRL2012}, which implies entanglement among atoms~\cite{GuhnePR2009} and/or atomic components~\cite{FernholzPRL2008}, has been demonstrated by quantum non-demolition measurement~\cite{MitchellNJP2012,SewellNP2013} including large degrees of squeezing using cavity enhancement~\cite{BohnetNPhot2014,HostenN2016}. 
Use of conditional spin-squeezed states has been demonstrated in magnetometry~\cite{SewellPRL2012} and clock operation~\cite{AppelPNAS2009, LerouxPRL2010, HostenN2016}. 
These works employ a measure-evolve-measure (MEM) sequence, in which a first quantum non-demolition (QND) measurement produces a state with reduced projection noise, a period of free evolution accumulates signal, and a second QND measurement detects the change relative to the first measurement. 
This method exploits the coherence of the atomic system, allowing signal to accumulate prior to readout of the atomic state, which typically destroys coherence. 

Here we demonstrate a generalization of this method, to allow quantification of arbitrarily-shaped signal components, using the same resources of atomic spin squeezing and QND measurement. 
This broadens the scope of entanglement-enhanced sensing to include arbitrary time-varying signals, of which magnetic waveforms due to firings of single neurons~\cite{JensenSR2016} and event-related fields in magnetoencephalography~\cite{SanderBOE2012} are notable examples. 
While rapid sampling of the same signals can also detect waveforms, it reduces $\tau$ and thus the sensitivity of the measurement. 
Our method includes radio-frequency (\RF) magnetometry as a special case, and we demonstrate entanglement-enhanced detection of radio fields in a compact, high-sensitivity \RF~magnetometer, with a sensitivity-volume figure of merit comparable to the best demonstrated instruments.

\PRLsection{Principle of the method} 
We consider an ensemble of atoms, described by a polarization ${\bf F}$, precessing in response to a magnetic field ${\bf B}(t) = \textbf{y} B_y(t) + \textbf{x} B_x(t)$, with $|B_y| \gg |B_x|$.
${\bf F}$ precesses about $y$ at an experimenter-controlled Larmor angular frequency $\Larmor(t) = \gamma B_y(t) + O(\gamma B_x^2/B_y) \approx \gamma B_y(t)$, driven transversally by the small unknown perturbation $B_x(t)$. 
The component $F_z$ is assumed accessible to QND measurement. The dynamics of the system are given by
\begin{equation}
\label{eq:dynamics}
\frac{d}{dt} {\bF}(t) = \gamma \bF(t) \times \bB(t) 
\end{equation}
where $\gamma$ is the gyromagnetic ratio of the atomic state. 
As shown in the \SuppInfo, for an initial atomic polarization oriented along $+y$ the evolution of the measurable spin component $F_z(t)$ can be expressed as
\begin{eqnarray}
\label{eq:sensing}
\fz(t) & = & \fz(0) \cos \Theta (t) +\fx(0) \sin \Theta (t) 
\\ & & 
+ \gamma \langle F_y(0) \rangle \int_0^t dt' \, \Bx(t') \cos[ \Theta(t) - \Theta(t') ] 
\nonumber \\ & & 
+ O(\Bx)^2 + O(\gamma t \Bx \delta F_y ) \nonumber 
\end{eqnarray}
where $\Theta(t) \equiv \int_0^t dt' \, \Larmor(t')$ is the accumulated angle.
We note that the first line describes an operator relation, namely a rotation of the spin components $\fz, \fx$ about $y$, and contains the quantum noise associated with the spin variables. 
As with other squeezing-enhanced atomic measurements, the noise in $F_z(t)$ can be reduced by squeezing a linear combination of $F_z(0)$ and $F_x(0)$. In particular, QND measurement of $F_z(0)$ can squeeze this component, while subsequent measurements at times $\{t_i\}$, chosen so that $\Theta(t_i)=n\pi$, $n \in \mathbb{Z}$, excludes noise of $F_x$ from entering the measurement record, achieving back-action evasion~\cite{VasilakisPRL2011, VasilakisNP2015}. 

In contrast, the second line of Eq.~(\ref{eq:sensing}) has no operator content, and describes a noiseless displacement by an amount proportional to the integral of the waveform $B_x(t')$ multiplied by the \textit{pattern function} $\cos[ \Theta(t) - \Theta(t') ]$. 
This describes a coherent build-up of the signal component matching the pattern function, and a cancellation of signal components orthogonal to it~\cite{deIcazaPRA2014}. 
By proper choice of $\Theta(t)$, the pattern function can be made to take on any functional form bounded by $\pm 1$. 
The third line of Eq.~(\ref{eq:sensing}) is negligible for the interesting case of weak signals and large atom number.

\begin{figure}[t]
\includegraphics[width=\columnwidth]{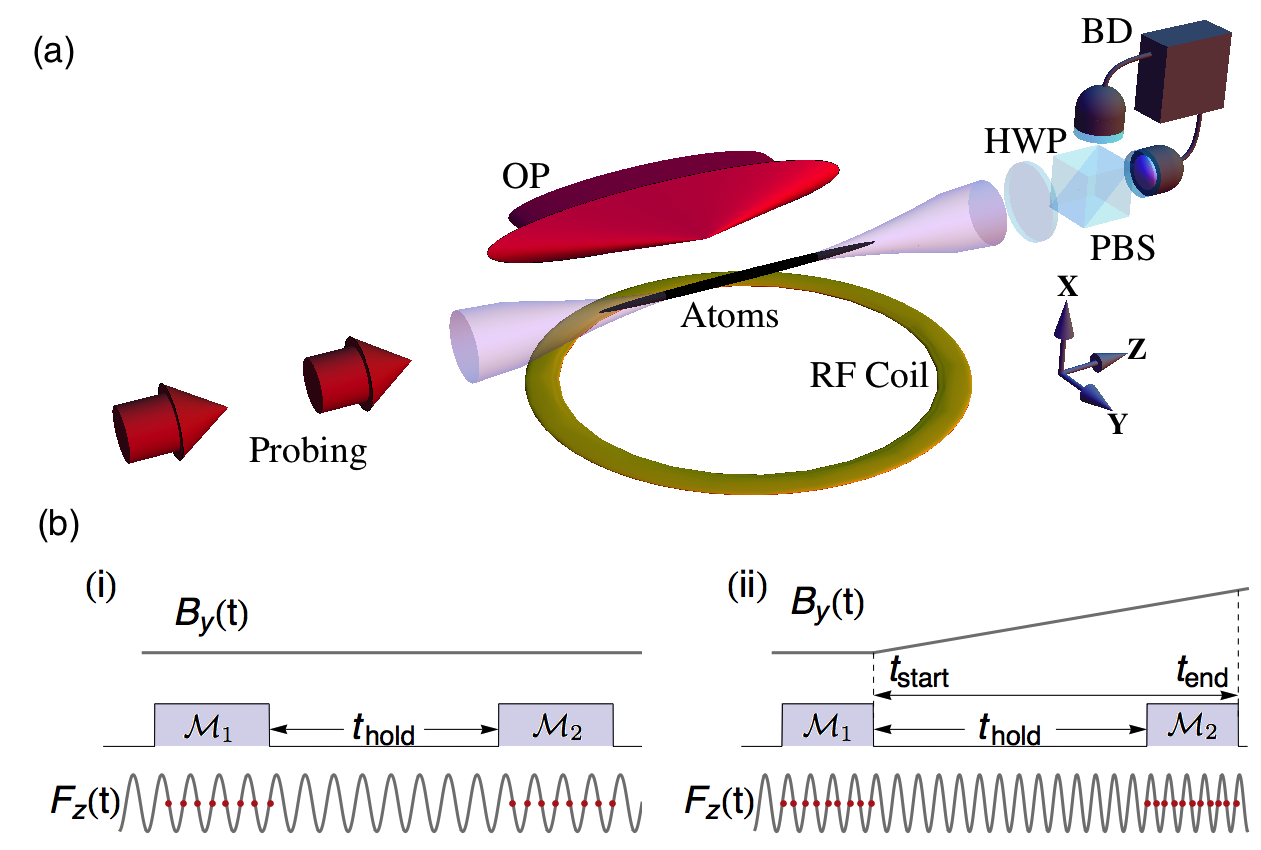}
\caption{
(a) Experimental geometry (not to scale), showing dipole-trapped atoms, on-axis Faraday rotation probe pulses, transverse optical pumping (OP), and polarimeter elements: 
half-wave plate (HWP), polarizing beamsplitter (PBS), and balanced detector (BD). Radio-frequency (\RF) magnetic fields in the $\hat{x}$ direction are produced by a coil, while a DC field of variable strength (not shown) is applied along $\hat{y}$. 
(b) Measure-evolve-measure sequence and illustration of back-action evading measurement of precessing spins experiencing, for (i) fixed $\Larmor$ or (ii) chirped $\Larmor(t)$. 
Probe times $t_i$ are illustrated by red dots, timed to give precession angle $\Theta(t_i) = n \pi$ for integer $n$.
\label{fig:Scheme}}
\end{figure}

\PRLsection{Experimental technique}
Our experimental apparatus, illustrated in Fig.~\ref{fig:Scheme} (a), is described in detail in~\cite{KubasikPRA2009}. 
Briefly, we trap up to $1.5\times10^6$ $^{87}$Rb atoms in a weakly focused single beam optical dipole trap. 
The atoms are laser cooled to a temperature of \SI{16}{\micro \kelvin} and optically pumped into the $f=1$ hyperfine ground state. 
A bias magnetic field along $y$ is generated with coils in a near-Helmholtz geometry fed by a programmable current source, and monitored using the atoms as an in-situ DC vector magnetometer~\cite{BehboodAPL2013}. 
An \RF~field along $x$ is produced with a low-inductance coil and an arbitrary waveform generator. 
Optical pumping (OP) along the $y$ direction is used to produce an initial atomic polarization, with an efficiency of $\sim 96\%$, as measured by Faraday rotation~\cite{KoschorreckPRL2010a, SupplementaryInformation}.

A non-destructive measurement of the atomic state is made using a train of \SI{600}{\nano\second} duration pulses of linearly polarized light \SI{700}{MHz} red detuned from the $f=1\rightarrow f'=0$ transition on the $\rm{D}_2$ line. 
The interaction between the atoms and the probe pulses is given by the effective Hamiltonian $\tau_{\rm{pulse}}\hat{H}_{\rm{eff}}=g_1\Fz \sz$ where the $g_1$ is a coupling constant depending on the probe beam geometry and detuning and $\tau_{\rm{pulse}}$ is the pulse duration, operators $F_i$ describe the atomic spins, and $S_i$ the optical polarization~\cite{SupplementaryInformation}.
Light pulses propagating along the trap axis experience a polarization rotation $\Syout = \Syin \cos \phi + \Sxin \sin \phi$, where $S_i^{({\rm in/out})}$ are Stokes operators before/after passing the atoms~\cite{KubasikPRA2009} and $\phi = g_1 F_z$ is the Poincar\'{e}-sphere rotation angle.
$\Syout$ is detected with a shot-noise-limited balanced polarimeter and $\Sxin = \NL/2$ is measured by splitting a constant fraction of the input light to a reference detector before the atoms. 
$g_1$ is calibrated by independent measurement~\cite{SupplementaryInformation}.

\begin{figure}[t]
\includegraphics[width=0.9 \columnwidth]{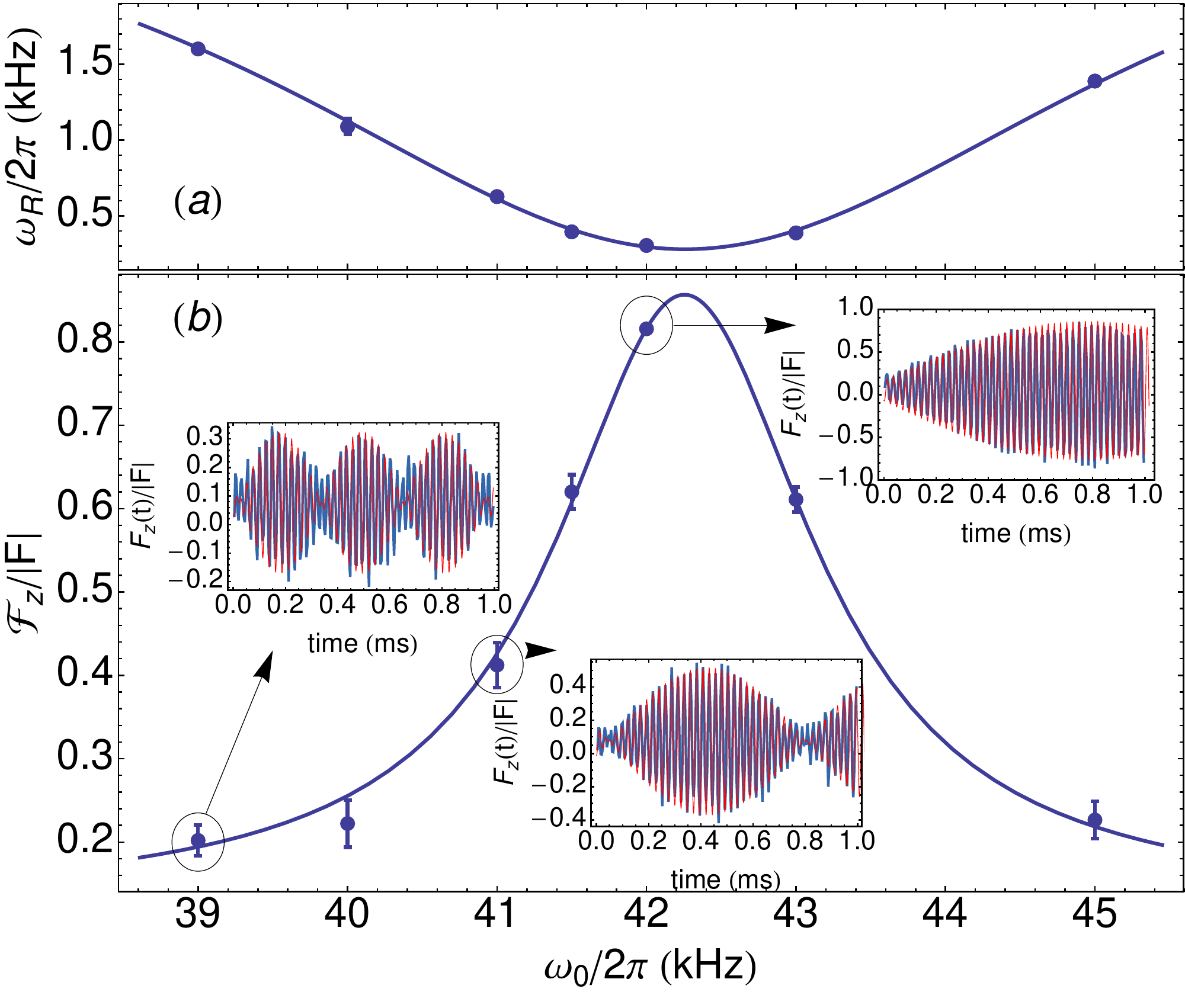}\\
\caption{
Characterization of the field-atom coupling. 
From an initially $+y$-polarized state, the ensemble spin is driven with continuous \RF~excitation $B_x(t)$ and constant bias field $\By$.
Insets show representative traces of atomic polarization $\fz$ versus time, normalized by the total atomic polarization $|{\bf F}|$. 
Blue curves show data and red curves show fits using Eq.~\eqref{eq:dynamics}. 
From such fits we obtain (a) Rabi oscillation frequencies $\omega_{\rm{R}}/2\pi$ and (b) maximal axial polarization ${\cal F}_z/|F|$, where ${\cal F}_z = \max |\Fz|$, versus applied \RF-signal frequency $\nu_{\rm{\RF}}$. 
In (a) and (b), solid blue lines show results of a Lorentzian fit, giving the resonance frequency $\omega_0/2\pi=$ \SI{42.26 (1)}{\kilo\hertz} and the \RF~field strength at the atoms, $a_{\rm{\RF}}$=\SI[parse-numbers=false]{0.60 (5)}{\milli G}.
Error bars show $\pm 1\sigma$ standard error of the mean.
\label{fig:Rabi}}
\end{figure}

\PRLsection{Calibration of response} 
The coherent response of the atoms is illustrated in Fig. \ref{fig:Rabi}. 
With a constant $B_y$ we apply \RF~ excitation of the form $\Bx(t)=a_{\rm{\RF}} \cos(\omega_0 t)$ and observe Rabi oscillation, i.e., sinusoidal oscillation of $\Fz$, amplitude modulated at the Rabi frequency as predicted by Eq. (\ref{eq:dynamics}).
We observe good experiment-theory agreement, and use the data as a calibration of $\By$ and $\Bx$. 

\begin{figure}[t]
\includegraphics[width= 0.9\columnwidth]{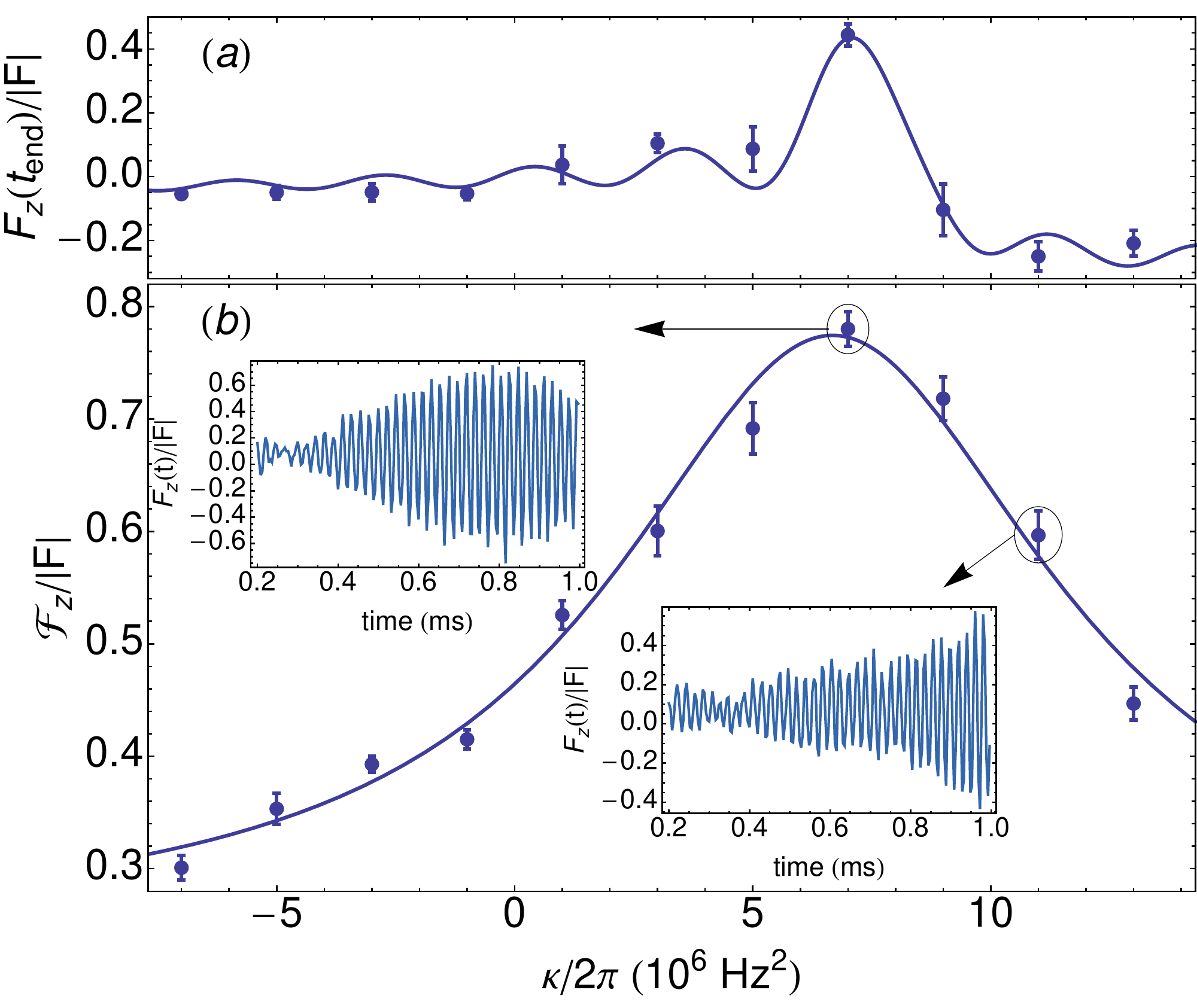}
\caption{
Response of the atomic polarization $\fz$ to \RF~waveforms with varying chirp $\kappa$. 
A ramped bias field $B_y(t)$ is used to produce a fixed Larmor frequency of $\Larmor(t)= 2 \pi \times \SI{42.2}{\kilo\hertz}$ until the end of $\Mone$, then rising linearly to $\Larmor(t)= 2 \pi \times \SI{47.5}{\kilo\hertz}$ over a period of $\SI{800}{\micro\second}$. 
During this time, an \RF~signal is applied with constant amplitude and frequency $\omega=  \omega_0 + 2 \kappa t$, and Faraday rotation probing is used to detect the axial polarization $\fz$.
Representative traces shown in the insets.
(a) Axial polarization $F_z(\Tend)/|F|$ at time $\Tend$, the end of the probe sequence. 
Solid line shows a numerical solution of Eq. (\ref{eq:dynamics}), and confirms the specificity for waveforms resembling the pattern function. 
(b) Maximal axial polarization ${\cal F}_z/|F|$ versus chirp of the \RF~excitation. 
Solid line is the result of a Lorentzian fit giving the resonant chirp $\kappa/2\pi= \SI{6.68(14)d6}{\hertz\squared}$.
Error bars show $\pm 1\sigma$ standard error of the mean.
\label{fig:transferChirp}}
\end{figure}

\PRLsection{Waveform detection}
Selective response to chirped waveforms is shown in Fig.~\ref{fig:transferChirp}. 
Using a ramped field $\By(t)$, we produce a Larmor frequency that sweeps linearly from $\Larmor\supone= 2 \pi \times \SI{42.2}{\kilo\hertz}$ to $\Larmor\suptwo= 2 \pi \times \SI{47.5}{\kilo\hertz}$ over $\SI{800}{\micro\second}$. 
This produces a ``chirped'' pattern function $\Theta(t)$, making the MEM sequence sensitive to $\Bx(t)$ signals with similar chirp, but insensitive to other waveforms, e.g. at constant frequency or with the opposite chirp. 
To confirm this selectivity, we use an arbitrary waveform generator to apply transverse fields of the form $\Bx(t)=a_{\rm{\RF}} \cos(\omega_0 t+ \kappa t^2)$, i.e., a linearly chirped waveform, in the time between $\Mone$ and $\Mtwo$. 

Fig.~\ref{fig:transferChirp} shows the resulting signal, i.e., the amplitude of the observed $\fz$ oscillation, as a function of the chirp $\kappa$. 
As expected, we observe a peak in the population transferred by the {\RF} drive when $\kappa$ matches the field ramp. 
Agreement with theory from Eq. (\ref{eq:dynamics}) is good, and variation in experimental signal is consistent with the independently-measured fluctuations of the magnetic field at the position of the atoms.

\PRLsection{Stroboscopic QND measurement}
To perform one QND measurement, $\Npulse$ probe pulses are sent through the atoms at intervals of one-half of the Larmor period, and experience a polarization rotation $\phi_n = g_1 \fz(t_n)$, where $n$ indexes the pulses.
Because of the inversion of $F_z$ between pulses, the $\phi_n$ can be aggregated as a single distributed measurement of $F_z$, quantified by
$\Phi \equiv {\Npulse}^{-1}\sum_{n=1}^{\Npulse}(-1)^{n-1}\phi_n.$
This multi-pulse probing has been shown to be a true QND measurement of the collective spin~\cite{SewellNP2013} with an uncertainty below the standard quantum limit for $F_z$~\cite{KoschorreckPRL2010a}. 

\PRLsection{Back-action evading \RF~sensing}
With these elements, we demonstrate back-action evasion in a MEM sequence to detect \RF~magnetic fields. 
We load the ODT with $\NA= \SI{1.5e6}{}$ atoms, measure the bias field as in~\cite{BehboodAPL2013} and then repeat the following MEM sequence 16 times: 
dispersive measurement of $\NA$ as in~\cite{KoschorreckPRL2010a}; optical pumping to produce full polarization along $+y$; QND measurement $\Mone$ with result $\Phi_1$; free evolution for time $\Thold=\SI{300}{\micro\second}$; and a second QND measurement $\Mtwo$ with result $\Phi_2$, as illustrated in Fig.~\ref{fig:Scheme} (b). 
During the hold time $\By$ is held constant, i.e. $\kappa = 0$, with $\Larmor=2 \pi \times \SI{50.16}{\kilo\hertz}$. The QND measurements $\Mone$ and $\Mtwo$ are made over \SI{200}{\micro\second} and contained \SI{4e8}{photons}, see \cite{SupplementaryInformation} for details.
The 16 repetitions of the MEM sequence allow us to vary $\NA$ since atoms are lost from the trap during optical pumping.
We repeat the full sequence 463 times to collect statistics.

\begin{figure}[t]
\includegraphics[width=\columnwidth]{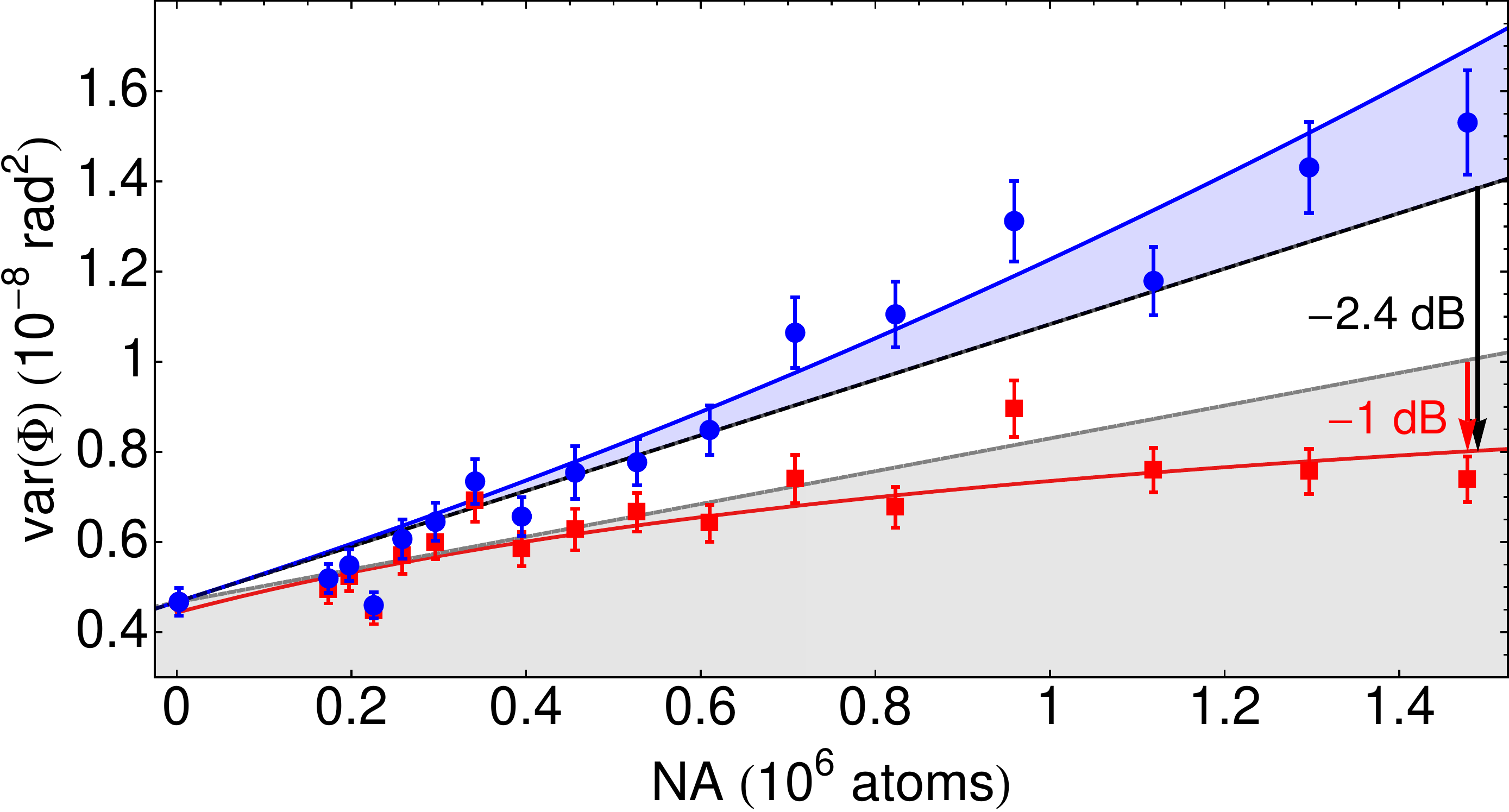}
\caption{
Atomic noise scaling of the stroboscopic QND measurement as a function of $\NA$ for constant bias field. 
Black dashed line shows calculated atomic projection noise for a coherent spin state (CSS) with $\var(\Phi)=\tilde{g}_1^2\NA/2$ plus readout noise (optical shot noise). 
Blue solid line is a quadratic fit to $\Phi_2$ (blue circles) using Eq.~(\ref{eq:variance}). 
Blue shaded region shows atomic technical noise. 
The red squares indicate the conditional variance $\mathrm{var}(\Phi_2|\Phi_1)$ as a function of $\NA$, and the solid red line quadratic fit to it. 
The dash-dotted line is the equivalent CSS projection noise, reduced by the loss of atomic coherence during $\Mone$. 
Gray shaded area shows region of metrologically-relevant spin squeezing. 
Error bars represent $\pm 1\sigma$ standard error.
\label{fig:Scaling}}
\end{figure}

\PRLsection{Projection noise level}
Fig.~\ref{fig:Scaling} shows the measured variance of $\Phi_2$ and the conditional variance $\var(\Phi_2|\Phi_1)$ as a function of the number of atoms in the trap. 
A linear measurement will show a variance that is quadratic in $\NA$~\cite{AppelPNAS2009}:
\begin{equation}
	\mathrm{var} (\Phi) =\mathrm{var}_0(\Phi) + \tilde{g}^2_1 \frac{1}{2} \alpha {\NA} + a_2 \NA^2
	\label{eq:variance}
\end{equation}
where $\mathrm{var}_0(\Phi)$ is the readout noise, quantified by repeating the measurement without atoms in the trap, $a_2 \NA^2$ is atomic technical noise associated with, e.g., fluctuations in state preparation, and the term $\propto \NA$ corresponds to atomic projection noise. 
The factor $1/2$ describes the $F_z$ variance of an $f=1$ atom polarized along $y$, and $\alpha$ accounts for the net noise reduction due to off-resonant scattering of probe photons. 
This scattering both reduces the number of noise-contributing atoms by pumping some into the far-off-resonance $f=2$ ground state, and adds noise as some atoms return to the $f=1$ state with randomized polarizations. 
We compute the evolution of the state using the covariance matrix methods reported in~\cite{ColangeloNJP2013} to find $\alpha = 0.96$.
The coupling $\tilde{g}^2_1$ is different from $g_1$ due to inhomogeneous atom-field coupling~\cite{VuleticPRA2015, AppelPNAS2009}.
A fit of Eq. (\ref{eq:variance}) to the data finds $\tilde{g}^2_1=1.2(2)\times 10^{-14}~\textrm{rad}^2$, determining the projection-noise level. 

\PRLsection{Squeezing}
To study the generation of squeezing we look at the correlation between $\Mone$ and $\Mtwo$. 
The first train of QND-pulses redistributes the noise to the non-measured component and $\Mtwo$ is used to evaluate its variance conditioned on the first measurement. 
The measurement noise reduction is quantified by $\mathrm{var}(\Phi_2|\Phi_1)=\mathrm{var}(\Phi_2-\chi \Phi_1)$, where $\chi=\mathrm{cov}(\Phi_1, \Phi_2)/\mathrm{var}(\Phi_1)>0$ describes the correlation between $\Phi_1$ and $\Phi_2$. 
As seen in Fig.~\ref{fig:Scaling}, the conditional variance is \SI[parse-numbers=false]{2.4 (2)}{dB} below the projection noise level.

\newcommand{\supMone}{^{(\Mone)}}
\newcommand{\supMtwo}{^{(\Mtwo)}}

Metrological improvement is quantified by the Wineland criterion~\cite{WinelandPRA1992} which takes into account the coherence loss and the noise reduction of the measured state: 
\begin{equation}
	\xi_m^2=\frac{1}{\eta^2}\frac{\mathrm{var}(\Phi_2|\Phi_1)}{\mathrm{var}\Phi_2}
\end{equation}
where $\xi_m^2<1$ indicates metrological advantage and $\eta$ accounts for the loss of coherence of the spin-squeezed state relative to the input coherent spin state.
The coherence after the first measurement is $\fy\supMone=\eta\fy$, where $\eta\equiv (1-\eta_{\rm{sc}})(1-\eta_{\rm{field}})$ and $\eta_{\rm{sc}}=0.11$ and $\eta_{\rm{field}}=0.04$ are independently measured coherence loss due to probe scattering and field inhomogeneities, respectively, during $\Mone$.
We find the metrological improvement due to squeezing $\xi_m^2=0.79(5)$, or \SI[parse-numbers=false]{1.0 (2)}{dB}.

\PRLsection{Entanglement-assisted waveform detection} 
To detect chirped waveform components beyond the projection-noise level, we repeat the above measurement and analysis using a chirped waveform with $\kappa = 2\pi\times\SI{5.78e6}{\hertz\squared}$, i.e., we ramp $\By(t)$ to produce a chirped pattern function during $\Thold=\SI{600}{\micro\second}$. The probe frequencies are matched to the Larmor precession frequency during $\Mone$ and $\Mtwo$ with \SI{2.9e8}{photons}, details in \cite{SupplementaryInformation}. 
For this experiment, $\NA = \SI{1.3e6}{}$.
We repeat the MEM sequence 21 times per trap loading to vary $\NA$, and the experiment 82 times to gather statistics.
Analyzed as above, we find \SI{1.5(3)}{dB} of noise reduction with \SI{0.8(3)}{dB} of metrological advantage. 
The reduced squeezing is due to the smaller $\NA$, smaller total photon number, and technical noise accumulated during the longer $\Thold$.

\PRLsection{Magnetic sensitivity} 
The sensitivity to the {\RF} drive amplitude $a_{\rm{\RF}}$ is $\delta B_{\rm{\RF}} = \Delta \phi/(| d\langle \phi\rangle /da_{\rm{\RF}}|)$ with $\phi$ being the observed Faraday rotation signals.
The signal accumulated during the $\Thold$ can be obtained by integration of Eq. \eqref{eq:sensing}, resulting in an enhanced single-shot sensitivity ~\cite{SupplementaryInformation}:
\begin{align}
\delta B_{\rm{\RF}}\sqrt{\Thold} &\simeq\frac{\Delta \phi_{\rm{cond}}}{g_1\eta_{\rm{hold}}\langle \Fy(0) \rangle}\frac{2}{\gamma ~\sqrt{\Thold}}
\end{align}
where $\langle \Fy(0) \rangle=\NA$, $\eta_{\rm{hold}}$ is the total coherence loss during $\Mone$ and $\Thold$, and $\Delta\phi_{\rm{cond}}^2=\mathrm{var}(\Phi_2|\Phi_1)$ is the conditional variance between the $\Phi_2$ and $\Phi_1$ optical signals, including the atomic noise and read-out noise.
Applied for the two {\RF} waveforms used, i.e., constant frequency and linearly chirped, we find sensitivities of 2.96$~\rm{pT/\sqrt{Hz}}$ and 3.36$~\rm{pT/\sqrt{Hz}}$, respectively. 
Focusing on the first, scaling the sensitivity by the volume of the atomic cloud, $V =1.43\times 10^{-5}\rm{cm^3}$, we find $\delta\rm{B}\sqrt{V}\approx 11.20~\rm{fT\sqrt{cm^3/Hz}}$. 
For comparison, the best alkali-vapor \RF~magnetometer~\cite{LeeAPL2006} in this frequency range showed a sensitivity of 0.24$~\rm{fT/\sqrt{Hz}}$ with $V=96~\rm{cm^3}$ or $\delta\rm{B}\sqrt{V}= 2.35~\rm{fT\sqrt{cm^3/Hz}}$. 
Thus the \RF~magnetometer demonstrated here has a volume-adjusted sensitivity comparable with the best existing instruments.

We have experimentally demonstrated detection of radio-frequency fields and arbitrarily-shaped radio-frequency waveform components beyond the projection noise limit, using stroboscopic back-action-evading measurements on magnetic atomic ensembles. The combination of QND measurements and stroboscopic probing in a measure-evolve-measure sequence gives this quantum sensing advantage, while also allowing full use of the system coherence, resulting in a sensitivity-volume figure of merit comparable to the best \RF~magnetometers at these frequencies.

\section*{Acknowledgements}
We thank C. Abell\'{a}n and W. Amaya for lending us the \RF~generator and S. Coop for useful feedback on the manuscript.
Work supported by MINECO/FEDER, MINECO projects MAQRO (Ref. FIS2015-68039-P), XPLICA (FIS2014-62181-EXP) and Severo Ochoa grant SEV-2015-0522, Catalan 2014-SGR-1295, by the European Union Project QUIC (grant agreement 641122), European Research Council project AQUMET (grant agreement 280169) and by Fundaci\'{o} Privada CELLEX.

%\bibliography{../bibliography/RFwWaveformBibv22}


\newpage
\begin{center}
\textbf{Supplementary Information}
\end{center}

\renewcommand{\theequation}{S.\arabic{equation}}
\renewcommand{\thefigure}{S.\arabic{figure}}

%\maketitle

\subsection{Operator definition}
We define the collective spin operation $\bF \equiv \sum_i \ff $, where $\ff$ is the spin of the $i$'th atom. The collective spin obeys commutation relations $[\Fx,\Fy]=i \Fz$, here and throughout we set $\hbar$=1. The probe pulses are described by the Stokes operator $\bss$ defined as $\Si\equiv\frac{1}{2}(\apd,\amd)\sigmai(\app,\am)^T$,where the $\sigmai$ are the Pauli matrices and $\apm$ are the annihilation operators for the $\sigma_{\pm}$ polarizations, which obey $[\Sx,\Sy]=i\Sz$ and cyclic permutations. The input pulses are fully $\Sx$ polarized, i.e., with $\ave{\Sx} = \NL/2$, $\ave{\Sy} =\ave{\Sz} = 0$ and $\Delta^2 S_i=\NL/4$, $i\in \{x,y,z\}$ where $\NL$ is the number of photons in the pulse. 

\subsection{State preparation}
\begin{figure}[htp]
\includegraphics[width=\columnwidth]{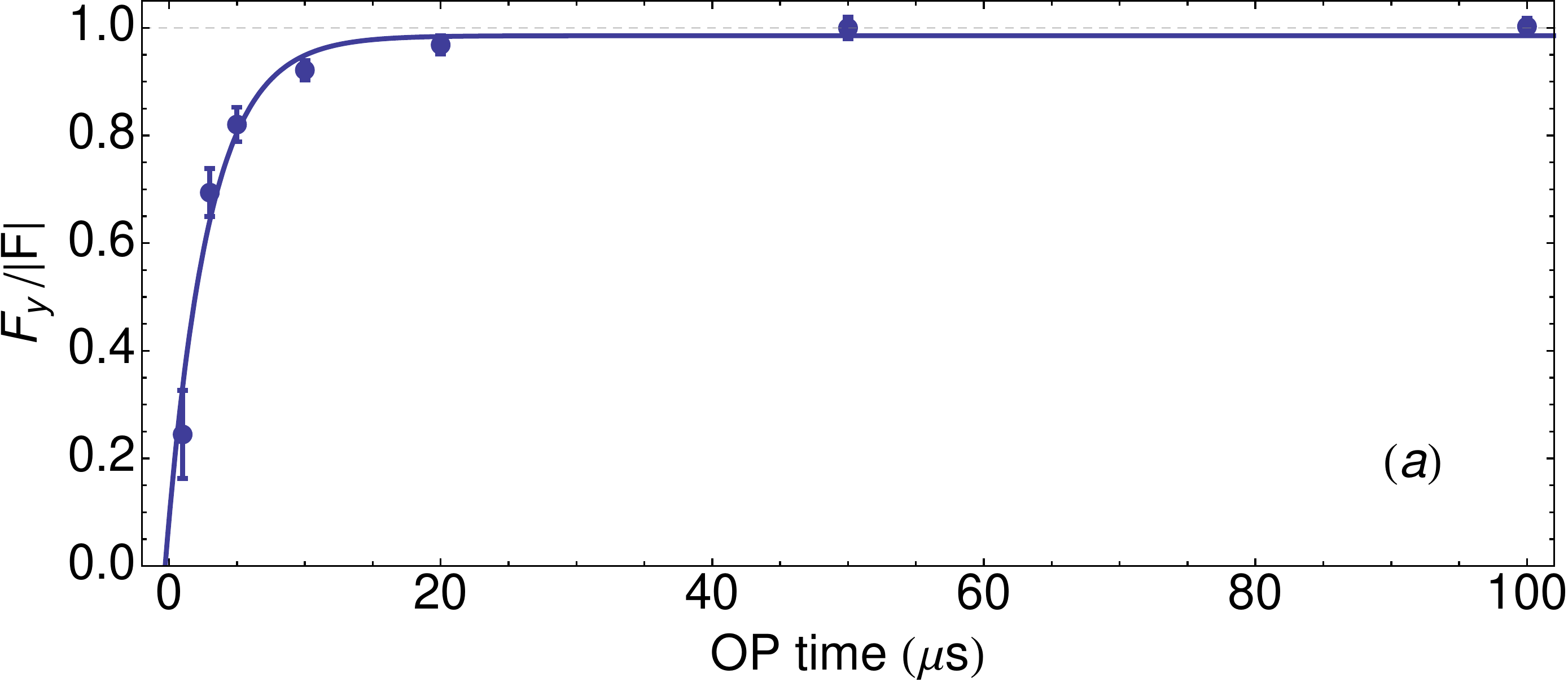}\\
\includegraphics[width=\columnwidth]{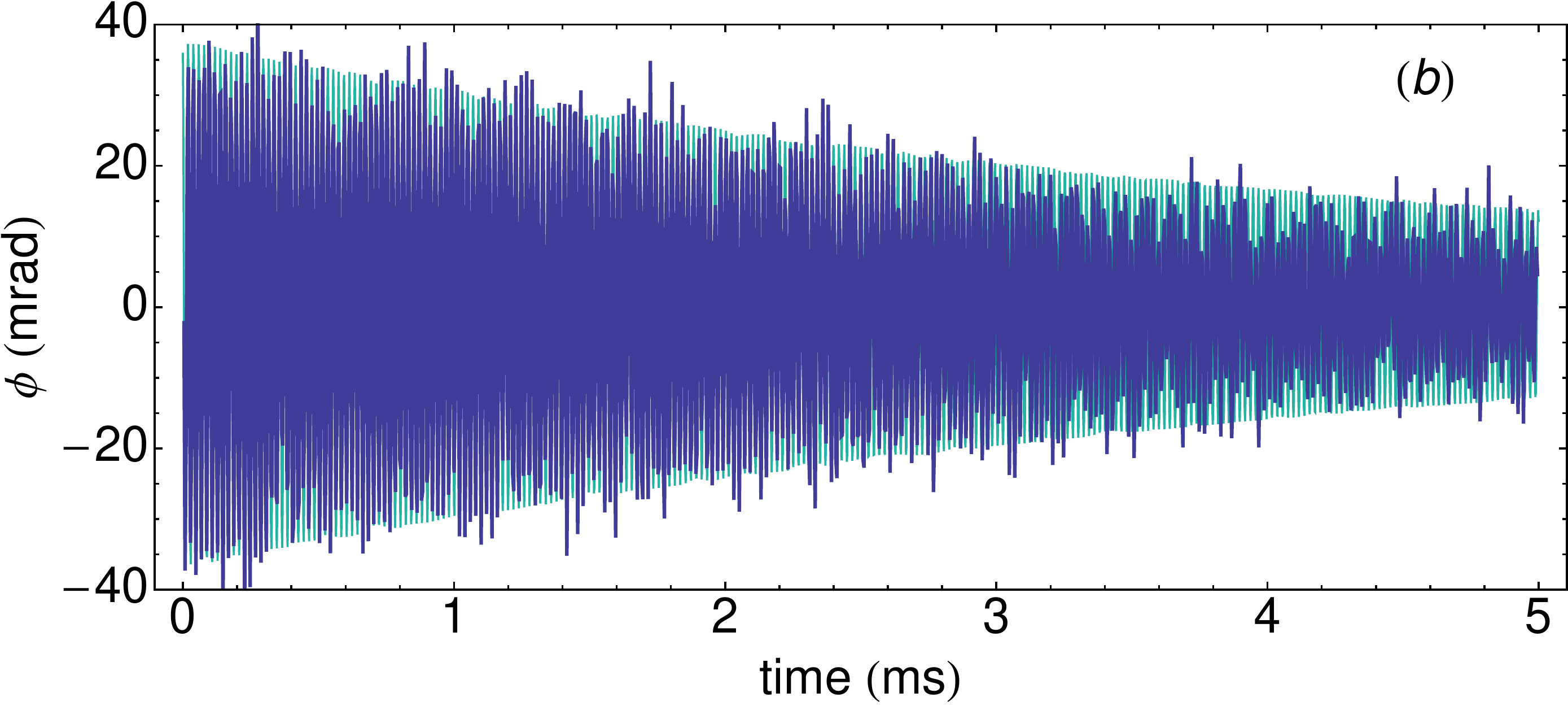} 
\caption{ (a) State preparation efficiency as a function of the optical pumping pulse length. The amplitude has been normalized to the total magnetization signal. The solid line is a fit the data using the curve $a(1- e^{-t/\tau})+c$ with fit outputs $a=0.97(4)$, $\tau=3.12(45)$ and $c=-0.01(6)$, from which we extract a pumping efficiency of $\Fy$ of $98\%$. Error bars represent $\pm 1\sigma$ standard error of the mean.
(b) Free induction decay (FID) measurement of the $\Fz$ polarized state precesing in a magnetic field $\By$, dark blue line. The pale green line is a fit with the function $\phi(t)=\beta+\alpha \cos(\Larmor t+\phi_0)\exp^{-t/T_2}$, giving $\beta=$\SI[parse-numbers=false]{0.4(2)}{\milli\radian}, $\alpha=$\SI[parse-numbers=false]{37.3(6)}{\milli\radian}, $\Larmor=2\pi\times$\SI[parse-numbers=false]{50.35(5)}{\kilo\hertz} and $T_2=$\SI[parse-numbers=false]{4.8 (3)}{ms}.}
\label{fig:calibrationPlots}
\end{figure}

The atoms are polarized so that $\langle{\bf F}\rangle$ is oriented along $+y$ via optical pumping under a parallel bias field $\By$.
We use a single \SI{50}{\micro\second} long pulse of circularly polarized light resonant with the $f=1\rightarrow f'=1$ transition of the D$_2$ line and propagating along the $y$-axis, and illuminate the atoms with repumper light resonant with the $f=2\rightarrow f=2'$ transition to prevent accumulation of atoms in the $f=2$ hyperfine level. 
To measure the atomic polarization, the atoms are subsequently rotated into $\Fz$ by slowly rotating the bias field from $\By$ to $\Bz$, and then measured with the Faraday probe. 
The amplitude of the transferred atoms is compared to the signal from an ensemble directly polarized by on axis optical pumping in a $\Bz$ field~\cite{KoschorreckPRL2010a}, resulting in a input polarized atomic ensemble with $\ave{\Fy}\simeq\NA$, Fig.~\ref{fig:calibrationPlots} (a). 

\subsection{Magnetic field measurement}
We make use of the atoms as an in-situ DC vector magnetometer to measure the applied magnetic field as described in~\cite{BehboodAPL2013}. We prepare an $\langle{\bf F}\rangle$ along $z$ under an orthogonal bias field $\By$ via optical pumping with circular polarized light propagating along the trap axis, and observe the free induction decay signal (FID) of the resulting Larmor precession using the Faraday probe. 
In order to observe the pure magnetic dephasing of the spins we probe semi-continuously with a reduced $\NL<10^6$ and large detuning, $\Delta=$\SI{1.5}{\giga\hertz}, to minimize probe scattering. 
We fit the FID traces with the function $\phi(t)=\beta+\alpha \cos(\Larmor t+\phi_0)e^{-t/T_2}$ and extract the Larmor angular frequency $\Larmor$ and the spin coherence time $T_2$. 
We minimize the amplitude of $\beta$ by compensating homogeneous field along $x$ and $z$ with coils in a near-Helmholtz geometry and cancel field gradients along the length of the trap by running current in opposite direction through pairs of wires transversal to the trap. The field is optimized in an iterative routine resulting in typical $\Ttwo=$\SI[parse-numbers=false]{4.8(3)}{\milli\second} and fields 99.1\% along the $y$-axis. A typical optimized FID signal is shown in Fig.~\ref{fig:calibrationPlots} (b).

\subsection{Rabi Flopping Calibration}
Adding an oscillatory (sinusoidal) driving field perpendicular to a static bias field induces Rabi flops between the magnetic sub-levels of the atoms in the hyperfine levels. 
When the frequency of the RF-signal matches the atomic energy splitting the flopping frequency is minimal and the population transferred by the RF maximal. 
As for the magnetic field measurement, in order to to minimize probe scattering the atoms are probed semi-continuously with a reduced $\NL$ and $\Delta=$\SI{1.5}{\giga\hertz}. 
The traces are fitted with the function 
\begin{equation}
\label{eq:fitRabi}
\Fz (t)=a \cos{(\Larmor t+\phi_L)}\cos{(\omega_{\rm{R}} t+\phi_R)}
\end{equation} 
where $\Larmor$ is the Larmor angular frequency and $\omega_{\rm{R}}$ is the Rabi flopping frequency. 
Fitting the maximal axial polarization $\mathcal{F}_z/|\Fz|$ as a function of the frequency of the RF excitation with a Lorenztian~\cite{PakePR1948} we determine the resonance frequency to be \SI[parse-numbers=false]{42.26 (1)}{\kilo\hertz} and estimate the strength of the RF-field coupled to the atoms to be $a=$\SI[parse-numbers=false]{0.60(5)}{\milli G}. 

\subsection{Ramped bias field}
\begin{figure}[htp]
\includegraphics[width=\columnwidth]{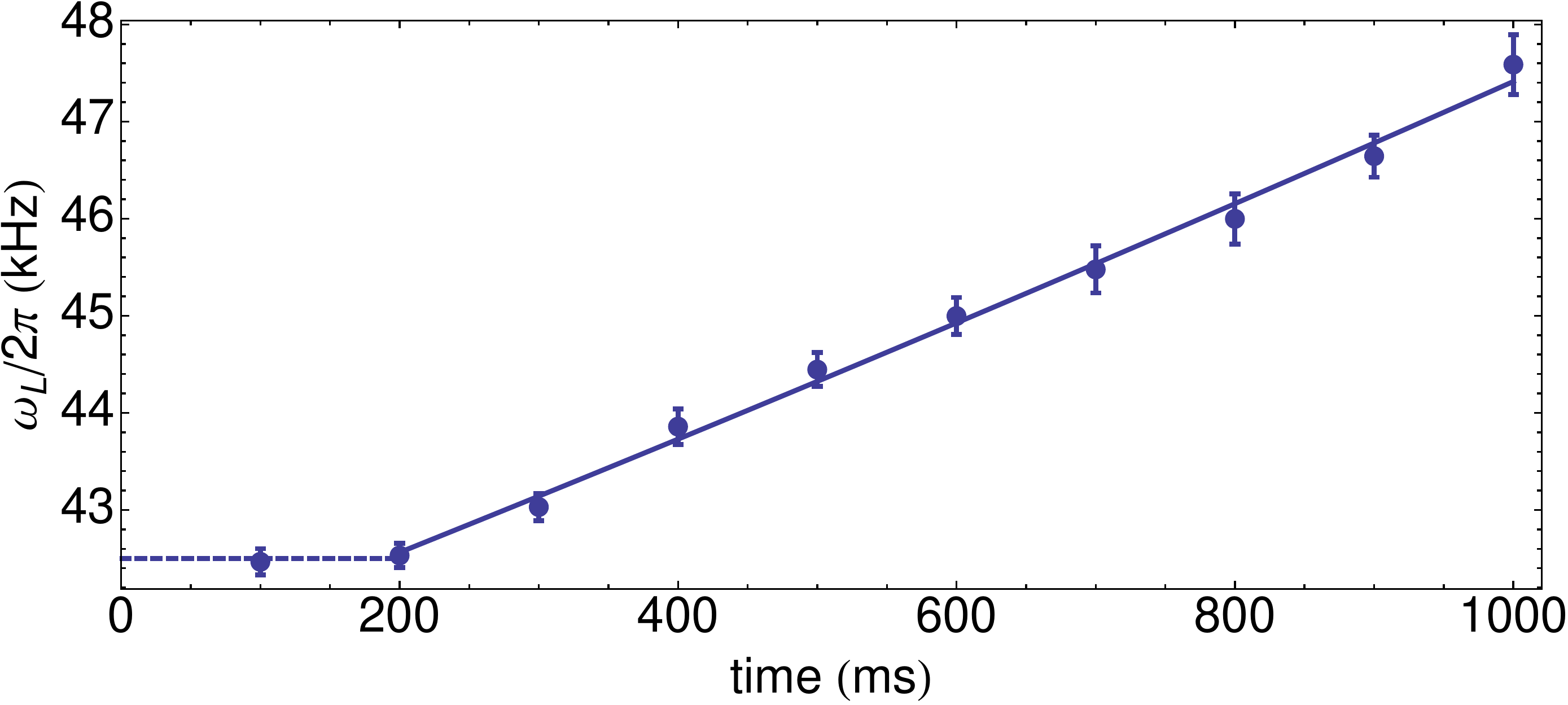}
\caption{
Change in Larmor frequency $\Larmor$ as a function of hold time. 
Solid line is a linear fit with slope $2\pi\times$ \SI[parse-numbers=false]{5.6 (1)\times 10^6}{\hertz\squared}. 
Broken line is the mean $\Larmor$ before the current of the $y$-coils is increased. 
Error bars represent $\pm 1\sigma$ standard error of the mean}.
\label{fig:frequencyChirp}
\end{figure}

To characterize the ramp of the bias field, we fit the FID signal with a chirped function of the form $\phi(t)=\beta+\alpha \cos(\Larmor t+\kappa t^2 +\phi_0)e^{-t/T_2}$. 
We find $\kappa=2\pi\times $\SI[parse-numbers=false]{5.6 (1)\times 10^6}{\hertz\squared} and $\Ttwo=$\SI[parse-numbers=false]{4.5 (4)}{\milli\second}. To confirm the linearity of the chirp we divide the FID-signal in \SI{100}{\micro\second} long segments and fit them individually with a function $\tilde\phi(t)=\beta'+\alpha' \cos(\omega_{\rm{L}}^{(i)} t+\phi_0')$, where $\omega_{\rm{L}}^{(i)}$ is the Larmor frequency of the $i$-th segment. 
A quadratic fit to the fit outputs for $\omega_{\rm{L}}^{(i)}$ confirms the linearity of the ramp as the quadratic term in negligible, $a_2 t/a_1=5.8\times 10^{-8}$, where $t=$\SI{800}{\micro\second} is the time during the ramp is on, see Fig.~\ref{fig:frequencyChirp}.

\subsection{Stroboscopic Probing}
We probe the atoms via off-resonant paramagnetic Faraday-rotation using $\tau_{\rm{pulse}}=$\SI{600}{\nano\second} duration pulses of linearly polarized light with a detuning of \SI{700}{MHz} to the red of the \rb D$_2$ line, and sent sent every half Larmor period, $\probePer=\Larmor/\pi$.
For the experiment with constant bias field during the hold time, $\Larmor=2 \pi \times \SI{50.16}{\kilo\hertz}$ and the QND measurements $\Mone$ and $\Mtwo$ are made at frequencies $\nuProbe\supone=\nuProbe\suptwo=\SI{101.01}{\kilo\hertz}$ over a time \SI{200}{\micro\second} giving a total of $\Npulse\supone = \Npulse\suptwo=20$ probe pulses, each with $\NL=2\times10^7$ photons. 
For the detection chirped waveform components the experiment was done ramping $\By(t)$ to produce a chirped pattern function with $\kappa = 2\pi\times\SI{5.78e6}{\hertz\squared}$ during $\Thold=\SI{600}{\micro\second}$.  The probe frequencies $\nuProbe\supone= \SI{84.75}{\kilo\hertz}$  $\nuProbe\suptwo=\SI{93.46}{\kilo\hertz}$ matched the Larmor precession frequency during $\Mone$ and $\Mtwo$ and contained $\Npulse\supone = 16$, $\Npulse\suptwo=18$, respectively, with $\NL=1.81\times10^7$ photons per pulse.
During a probe pulse the atomic spins rotate by an angle $\theta=\gamma~B~ \tau_{\rm{pulse}}$. For our parameters, $\tau_{\rm{pulse}}$ and $B=|\bB|\lesssim \SI{70}{\milli G}$ we find $\Theta\sim$ \SI{30}{\milli\radian}, so we can neglect the rotation of the spins during the probe pulse. 

\subsection{Volume estimation}
\label{section:volume}
We use absorption imaging to estimate the volume of the atomic cloud. Atoms originally in the $f=1$ hyperfine ground state are transferred into the $f=2$ hyperfine ground state by a \SI{100}{\micro\second} pulse of laser light tuned to the $f=1\rightarrow f'=2$ transition. A \SI{100}{\micro\second} pulse of circularly polarized light resonant to the $f=2\rightarrow f'=3$ transition cast a shadow on a CCD camera. To avoid spatially dependent light shifts the dipole trap is switched off during the process.
To estimate the radial extension of the atom we use the time-of-flight (TOF) technique~\cite{KubasikPRA2009}. Free, thermal and isotropic expansion of the atoms is described by
%The size of the ensemble is described by Free thermal, isotropic expansion of the atom cloud dictates that the cloud radius
\begin{equation}
\label{eq:expansion}
\omega_a(t)^2=\omega_a^2(0)+\frac{k_B T}{m}(t-t_0)^2
\end{equation}
where $k_B$ is Boltzmann constant and $m$ the atomic mass.
$\omega_a(0)$ is the initial size of the ensemble, $T$ its temperature and $t_0$ is a time delay for switching off the dipole trap, all 3 free parameters of the fit. 

\begin{figure}[htp]
\includegraphics[width=\columnwidth]{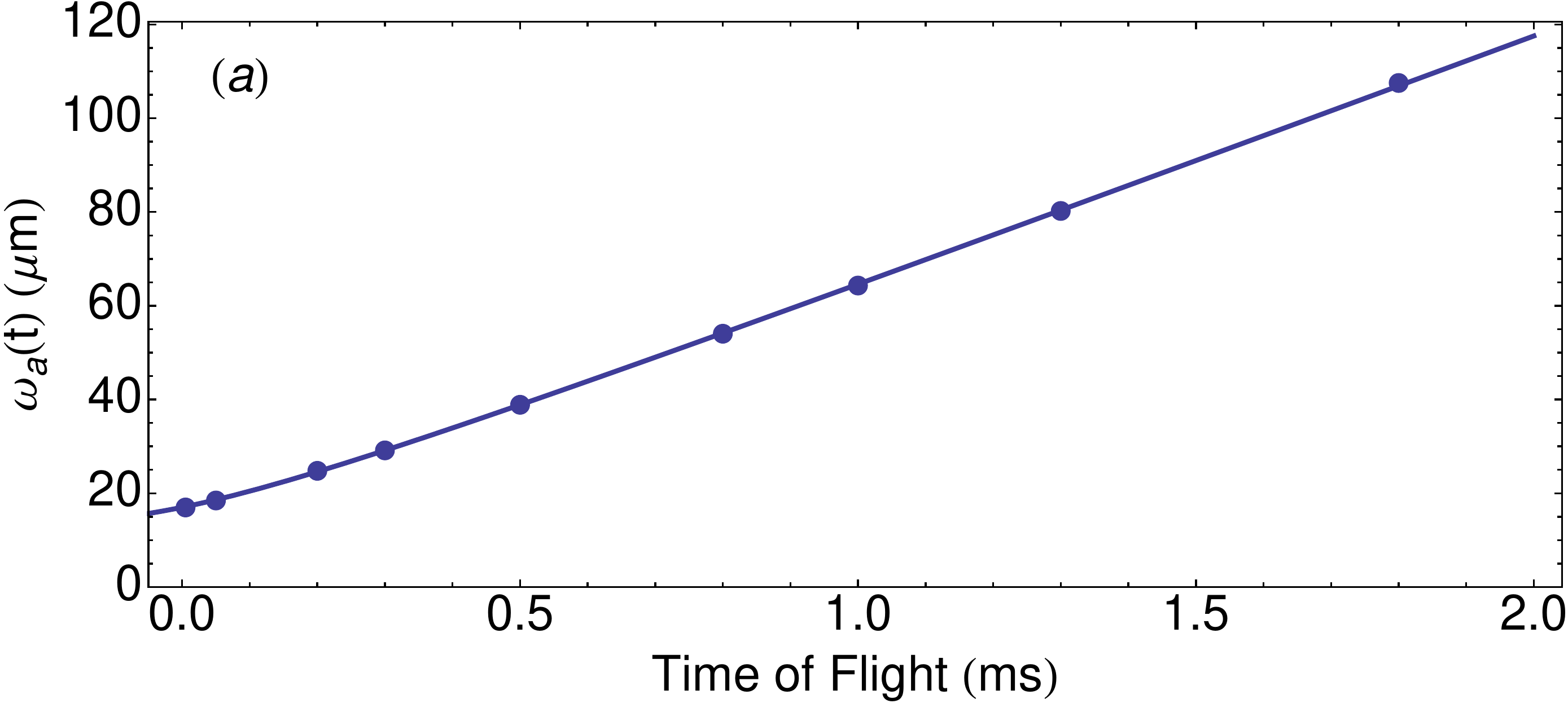}\\
\includegraphics[width=\columnwidth]{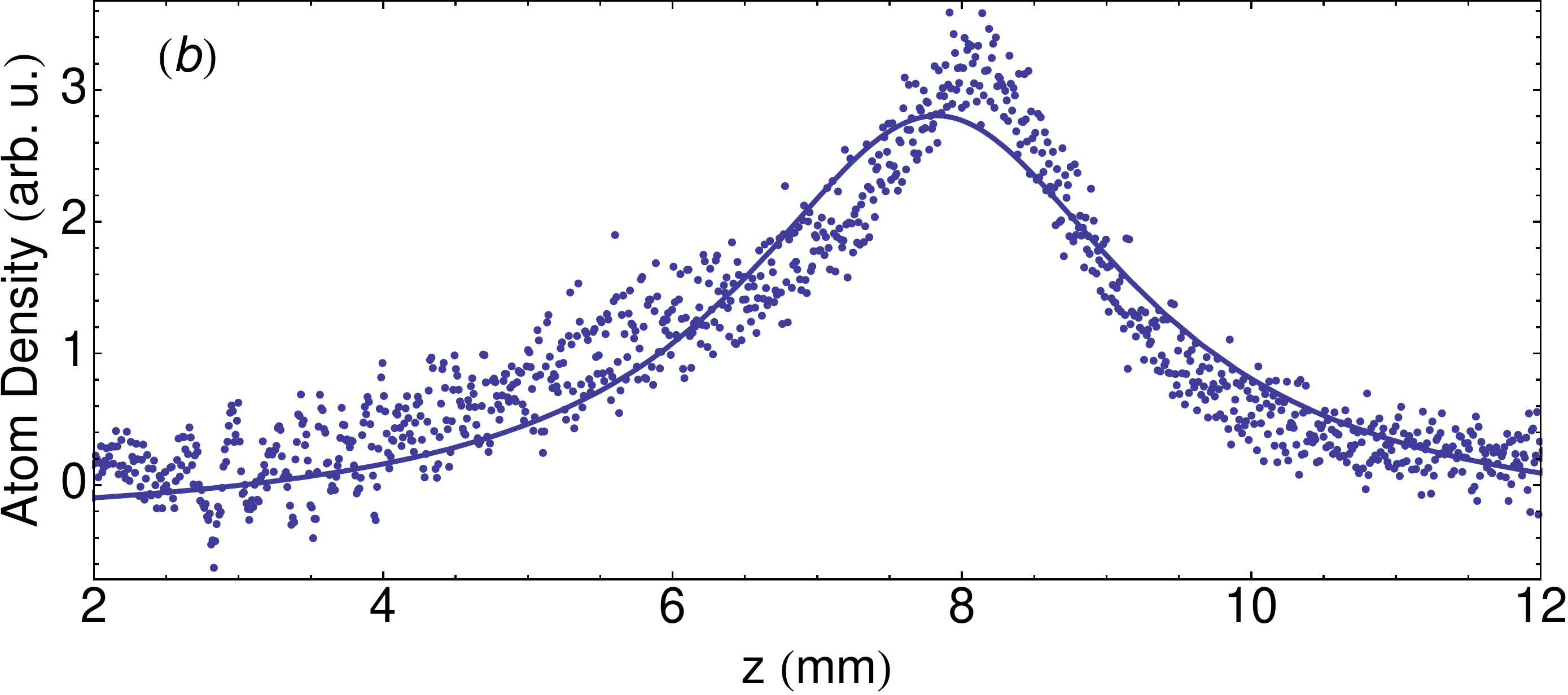} 
\caption{
(a) Characteristic transverse width of the expanding atomic cloud for different free falling times. 
Error bars would be smaller than the point and are not represented. 
Solid line is a fit using Eq.~\eqref{eq:expansion} used to determine the temperature of the atoms \SI{15.5 (1)}{\micro\kelvin} and the transverse initial size of the ensemble $\omega_a(0)=$\SI{14.2 (4)}{\micro\meter}.
(b) Axial shape of the atomic ensemble. 
The solid line is a fit with a Lorentzian function used to estimate the longitudinal FWHM $\rho_{long}=$\SI{3.28 (6)}{\milli\meter}.}
\label{fig:sizecloud}
\end{figure}

We determine the width of the radial profile by integrating each frame of the image along the longitudinal direction (the $z-$axis) and fit the resulting density profile with a Gaussian to get the center and the width of the atomic ensemble. 
The width of the atomic cloud $\omega_a$ is plotted on Fig.~\ref{fig:sizecloud} (a) as a function of time. 
We fit with Eq.~\eqref{eq:expansion} to find $\omega_a(0)=$\SI{14.2 (4)}{\micro\meter}, from which we determine the radial full width at half maximum (FWHM) of the atomic cloud $\rho_{radial}=2\sqrt{2~\rm{ln}2}~\omega_a(0)=$\SI{33 (1)}{\micro\meter}.
The temperature of the atomic sample is \SI{15.5 (1)}{\micro\kelvin}

The axial dimension of the trap is much longer than the spread of the ensemble during the TOF, resulting in $\rho_{axial}(t)\simeq\rho_{axial}(0)$.
The axial shape of the atomic ensemble is calculated by integrating the images along the transverse direction, Fig.~\ref{fig:sizecloud} (b). 
The black solid line is a fit with a Lorentzian of the form $L(z)=a\rho_{axial}/2((z-z_0)^2+(1/2\rho_{axial})^2)+b$, where $\rho_{axial}$ is the FWHM. 
We determine the axial FWHM atomic length $\rho_{axial}=$\SI{3.28 (6)}{\milli\meter}.

We approximate the atomic volume to an ellipsoid with semi-axis given by $\rho_{axial}$ and $\rho_{radial}$ and calculate its volume using $V=\frac{4\pi}{3}\rho_{radial}^2\rho_{axial}$=\SI{1.43e-5}{\centi\meter\cubed}. 

\newcommand{\supzero}{^{(0)}}

\subsection{Waveform component detection}

The dynamics of the spins are governed by the Heisenberg equations of motion (with $\hbar = 1$):
\begin{equation}
\label{eq:dynamics}
\frac{d}{dt} {F_i} = -i [F_i,H_0(t)+ H'(t)]
\end{equation}
where $H_0 = - \gamma {\bf F} \cdot {\bf B} = - \gamma F_y B_y(t)$, $\gamma$ is the gyromagnetic ratio for the $F=1$ ground hyperfine state, and the perturbation $H' = - \gamma F_x B_x(t)$ describes the RF drive. We use a Dyson series to solve the resulting system of differential equations. We define ${\bf F}^{(0)}(t)$ to be the solution to Eq.~\eqref{eq:dynamics} when $H' = 0$, i.e. ${F}\supzero_y(t) = {F}_y(0)$ and 
\begin{eqnarray}
\label{eq:transformation}
\left( 
\begin{array}{c}
\Fz\supzero(t) \\ \Fx\supzero(t)
\end{array}
\right) &=& \left( 
\begin{array}{cc}
\cos \Theta(t) & \sin \Theta(t) \\
-\sin \Theta(t) & \cos \Theta(t)
\end{array}
\right) \left( 
\begin{array}{c}
\Fz(0) \\ \Fx(0) 
\end{array}
\right) ~~
\end{eqnarray}
where $\Theta(t)\equiv \int_0^t dt' \, \omega_L(t')$, $\Larmor(t) \equiv \gamma B_y(t)$ is the accumulated angle.
We then use the well-known result~\cite{Sakurai}
\begin{equation}
\label{eq:eq_temporal}
F_i(t) = F_i^{(0)}(t) - i \int_0^t dt' \, [F_i^{(0)}(t), H'(t')] + O(H')^2
\end{equation}
which, in light of the definition of $H'(t')$, gives the signal
\begin{eqnarray}
\label{eq:sensing}
\Fz(t) & = & \cos \Theta (t) \Fz(0) + \sin \Theta (t) \Fx(0) 
\nonumber \\ & & 
+ \gamma \int_0^t dt' \, F_y(t') B_x(t') \cos[ \Theta(t) - \Theta(t') ] 
\nonumber \\ & & + O(B_x)^2 
\end{eqnarray}
We note that $F_y(t') = F_y(0) + O(B_x)^2$ and writing $\delta F_y^2 = \langle F_y^2 \rangle - \langle F_y \rangle^2 $, we find 
\begin{eqnarray}
\label{eq:sensing}
\Fz(t) & = & \cos \Theta (t) \Fz(0) + \sin \Theta (t) \Fx(0) 
\nonumber \\ & & 
+ \gamma \langle F_y(0) \rangle \int_0^t dt' \, F_y(t') B_x(t') \cos[ \Theta(t) - \Theta(t') ] 
\nonumber \\ & & + O(B_x)^2 + O(\gamma t B_x \delta F_y)^2. 
\end{eqnarray}
The term $O(B_x)^2$ describes contributions of higher order in $H'$, and can be neglected in the interesting scenario of weak signals. For our initial state of all atoms pumped into the $y$ direction, $\delta F_y \ll F_y$, making the last term negligible also.

Dropping the higher other terms and choosing the probing times such that $\Theta(t_i)=n\pi$ simplifies Eq.~\eqref{eq:sensing} further, to 
\begin{equation}
\label{eq:sensingsimple}
\Fz(t_n)=(-1)^n \left(\Fz(0) + \gamma \langle F_y(0) \rangle \int_0^{t_n} dt' \, \Bx(t') \cos\Theta(t')\right)
\end{equation}

\subsection{Field sensitivity}
The accumulated signal during the hold time $\Thold$ is obtained by integration of Eq.~\eqref{eq:sensingsimple}, where for the case of ramped field, $B_x(t') = a_{\rm{\RF}}\cos(\omega_{0} t'+\kappa t'^2)$, and $\Theta(t) = \omega_L t + \kappa_L t^2$, where $\kappa_L =(\omega_f-\omega_i)/\Delta t$ describes the change of the Larmor frequency caused by the ramp, applied for a time $\Delta t$. The radio-frequency case is included has a special case when $\kappa=\kappa_L=0$.

The solution to Eq.~\eqref{eq:sensingsimple} becomes
\begin{align}
\label{eq:FzRF}
	F_z(\Thold) &= \Fz(0) - \gamma \langle \Fy(0) \rangle \times \nonumber \\
	&\int_0^{\Thold} dt' \, a_{\rm{\RF}} \cos(\omega_{0} t'+\kappa t'^2) \cos(\omega_L t' + \kappa_L t'^2)
\nonumber \\
&= \Fz(0)+ \gamma a_{\rm{\RF}} \langle \Fy(0) \rangle I(\omega_L,\omega_{0},\kappa_L,\kappa, \Thold)
\end{align}
Evaluating the integral on resonance, i.e. with $\omega_0 = \omega_L$, and $\kappa_L=\kappa$. In cases of interest $\kappa \ll \omega_0$, giving the leading-order in $I(\omega_L,\omega_{0},\kappa_L,\kappa, \Thold)\simeq\Thold/2$, with an error smaller than $3\%$ for our parameters.

The sensitivity of a coherent spin state to the {\RF} drive amplitude $a_{\rm{\RF}}$ during a time $\Thold$ is
\begin{align}
\delta B_{\rm{\RF}}^{\rm{CSS}} &= \frac{\Delta \phi}{\left| d\langle \phi\rangle /da_{\rm{\RF}}\right|} \nonumber \\
&= \frac{\Delta \phi}{g_1\langle \Fy(0) \rangle}\frac{1}{\gamma ~I(\omega_L,\omega_{0},\kappa_L,\kappa,\Thold)} \nonumber \\
&\simeq\frac{\Delta \phi}{g_1\langle \Fy(0) \rangle}\frac{2}{\gamma ~\Thold}
\end{align}
where $\langle \Fy(0) \rangle=\NA$.  For the spin squeezed state 
\begin{align}
\delta B_{\rm{\RF}}^{\rm{SSS}} &\simeq\frac{\Delta \phi_{\rm{cond}}}{g_1\eta\langle \Fy(0) \rangle}\frac{2}{\gamma ~\Thold}
\end{align}
where $\eta$ is the coherence loss including $\Thold$ and the first measurement and $\Delta\phi_{\rm{cond}}^2=\mathrm{var}(\Phi_2|\Phi_1)$ is the conditional variance between the $\Phi_2$ and $\Phi_1$ optical signals, including the atomic noise and read-out noise.

\subsection{Statistics of probing inhomogeneously-coupled atoms}
We consider the statistics of the Faraday rotation measurements of an ensemble of $\NA$ atoms, described by individual spin operators ${\bf f}_i$. To define the SQL, we consider the atomic ensemble to be in a coherent spin state, with independent individual spins and fully polarized with $\ave{\Fy}\simeq\NA$. When the spatial structure of the probe beam is taken into account, the Faraday rotation is described by the input-output relation for the Stokes component $\Sy$
\begin{eqnarray}
\label{eq:FSQL}
\Sy^{\rm{(out)}} &=& \Sy^{\rm{(in)}} + \Sx^{\rm{(in)}} \sum_{i=1}^{\NA} g({\bf x}_i) \fz^{(i)} 
\end{eqnarray}
where $g({\bf x}_i)$ is the coupling strength for the $i$-th atom, proportional to the intensity at the location ${\bf x}_i$ of the atom. $\Sy^{\rm{(in)}}$ has zero mean and variance $\NL/2$, where $\NL$ si the mean photon number in the probe pulse $\Sx$. 
The atomic spin is polarized along $\Fy$ and orthogonal to the measured $\Fz$ direction. The rotation angle $\phi = \Sy^{\rm{(out)}}/\Sx^{\rm{(in)}}$ has statistics
\begin{eqnarray}
\langle \phi \rangle &=& \langle \fz \rangle \sum_{i=1}^{\NA} \left< g({\bf x}_i) \right>_{{\bf x}_i} 
\nonumber \\ 
& \equiv & \langle \fz \rangle \langle \NA \rangle g_1 \\
\mathrm{var}(\phi) &=& \mathrm{var}(\phi_0) + \mathrm{var}(\fz) \left< \sum_{i=1}^{\NA} g^2({\bf x}_i) \right>_{\NA,{\bf x}_i}
\nonumber \\ 
& \equiv & \mathrm{var}(\phi_0) + \mathrm{var}(\fz) \langle \NA \rangle \tilde{g}_1^2
\end{eqnarray}
where $\phi_0$ is the polarization angle of the input light, subject to shot-noise fluctuations and assumed independent of $\Fz$, and the angle brackets indicate an average over the number and positions of the atoms. For known $\langle \fz \rangle$ and $\mathrm{var}(\fz)$, measurements of $\langle \phi \rangle$ and $ \mathrm{var}(\phi)$ versus $\NA$ give the calibration factors $g_1$ and $\tilde{g}_1^2$. 

\subsubsection{Calibration of the $g_1$ factor}
\begin{figure}[htp]
\includegraphics[width=\columnwidth]{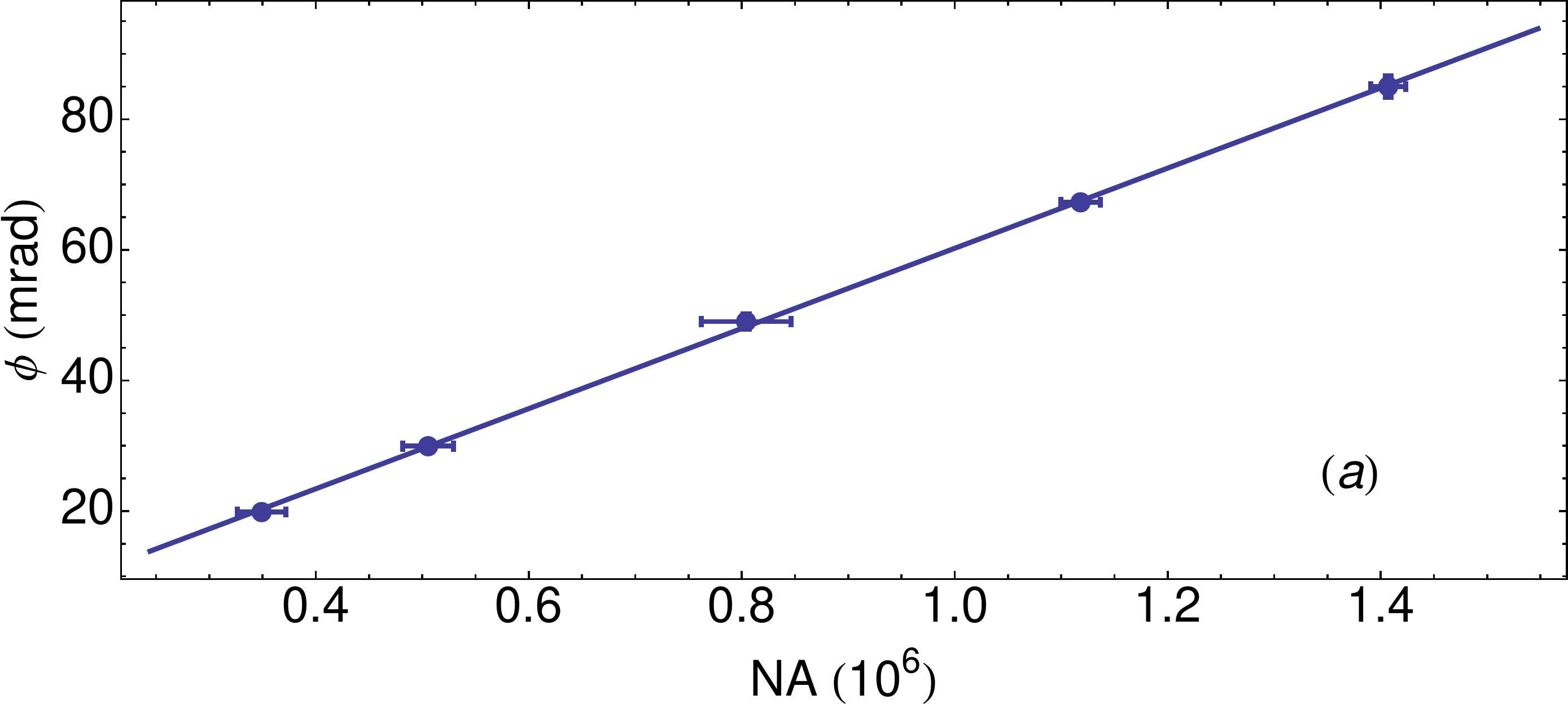}\\
\includegraphics[width=\columnwidth]{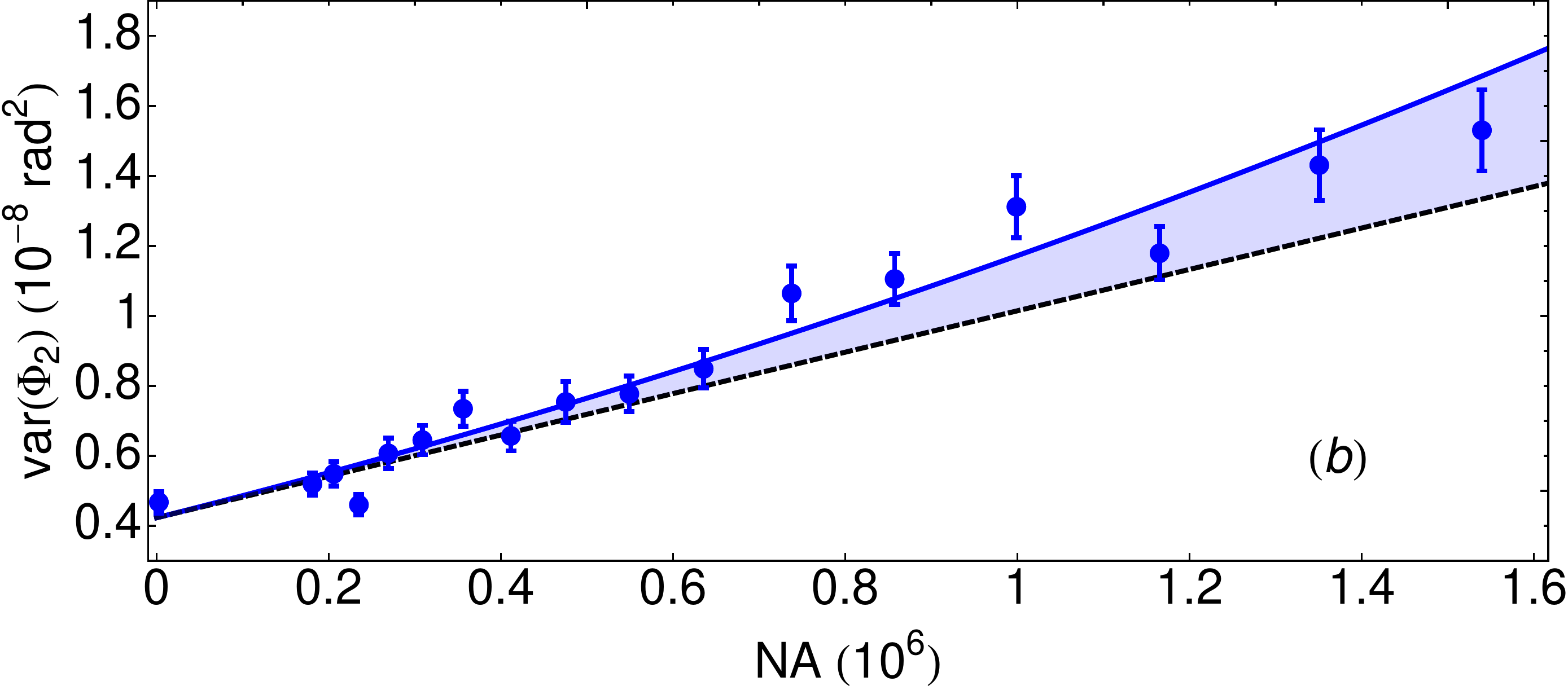} 
\caption{
(a) Calibration of mean Faraday rotation signal against input atom number $\NA$ measured via absorption imaging. Solid line, the fit curve $\phi= a_0 + g_1 \NA$ with values $g_1=6.16(9)\times 10^{-8}$~radian$\cdot$atom$^{-1}$ and $a_0=1.42(63)\times 10^{-3}$~radian. Error bars indicate $\pm 1\sigma$ statistical errors
(b) Calibration of the quantum noise limited Faraday rotation probing of spins. 
We plot the variance of $\Phi_2$ as a function of the number of atoms $\NA$ in an input coherent state polarized along $\Fy$. 
Solid curve is a fit using Eq.~\eqref{eq:variance}, we estimate $\mathrm{var}(\Phi_0)=4.2(2)\times 10^{-9}$, $\tilde{g}_1^2=1.2(2)\times 10^{-14}$ and $a_2=1.3(8)\times 10^{-21}$. 
The blue region represents the presence of technical noise in the input atomic state. 
The black dashed line indicates the expected atomic projection noise for an ideal CSS with $\mathrm{var} (\Phi) =\mathrm{var}(\Phi_0) + \tilde{g}_1^2 \frac{1}{2} \alpha {\NA}$. 
Error bars indicate $\pm 1\sigma$ standard error in the variance. 
\label{fig:g1calibrations}}
\end{figure}

We calibrate $\langle \phi \rangle$, the measured rotation angle of the dispersive atom number measurement (DAMN)~\cite{KoschorreckPRL2010a} against an independent estimate of $\NA$ made via absorption imaging.
The results are shown in Fig.~\ref{fig:g1calibrations} (a). 
For this experiment, the atoms are prepared in an $\Fz$-state with a single \SI{50}{\micro\second} duration circularly polarized optical pumping pulse on resonance with the $f=1\rightarrow f'=1$ transition of the D$_2$ line and propagating along the trap axis with an bias field $\Bz$=\SI{180}{mG} applied to fix the atomic polarization, and then probed with the Faraday probe. 
The number of atoms is estimated using absorption imaging in the same way as explained in \textit{Volume estimation}.  
We calculate the resonant interaction cross-section and take into account the finite observable optical depth. The statistical error in the absorption imaging is $<3\%$, including imaging noise and shot-to-shot trap loading variation.

\subsubsection{Calibration of the $\tilde{g}_1^2$ factor}

To measure $\tilde{g}^2_1$ we use the the stroboscopic QND data. The atoms are prepared into an $\Fy$-polarized state and then probed stroboscopically at twice the Larmor period, in such a way that the measured variable is $\pm \Fz$, evading back-action effects. If $\phi_n$ is the measured Faraday rotation angle for pulse $n$, and $\phi_0^{(n)}$ is the corresponding input angle, we can define the pulse-train-averaged rotation signal as 
\begin{eqnarray}
\phi &\equiv & \frac{1}{\Npulse}\sum_{n=1}^{\Npulse} (-1)^{n+1}\phi_{n} 
\end{eqnarray}
with variance
\begin{eqnarray}
\mathrm{var}(\phi) &\equiv & \mathrm{var}( \phi_0 ) + \tilde{g}_1^2 \sum_{n=1}^{\Npulse} \mathrm{var}(F_{z,n})
\end{eqnarray}
where $\phi_0 = \tfrac{1}{\Npulse}\sum_{n=1}^{\Npulse} \phi_0^{(n)}$, with zero mean and variance $\mathrm{var}(\phi_0) = (\Npulse \NL)^{-1}$, and ${F}_{z,n}$ is the value of $\Fz$ at the time of the $n$-th probe pulse. 

We compute the resulting evolution of the state using the covariance matrix methods described in~\cite{ColangeloNJP2013} and presented in the following section resulting in 
\begin{eqnarray}
\mathrm{var}\left( \frac{g_1}{\Npulse} \sum_{n=1}^{\Npulse} {F}_{z,n} \right) & = & \tilde{g}_1^2 \frac{1}{2} \alpha {\NA}
\end{eqnarray}
corresponding to the linear term in $\NA$ used in Eq.~(4) of the main text to determine the atomic quantum noise, i.e.,
\begin{equation}
	\mathrm{var} (\Phi) =\mathrm{var}(\Phi_0) + \tilde{g}_1^2 \frac{1}{2} \alpha {\NA} + a_2 \NA^2,
	\label{eq:variance}
\end{equation}
where the correction factor $\alpha=0.96$ accounts for decoherence and noise introduced into the atomic state due to off-resonant probe scattering during the QND measurement~\cite{ColangeloNJP2013}.

%\bibliography{../bibliography/rfBibSI}


\end{document}